\documentclass[amssymb,nofootinbib]{revtex4}
\usepackage{epsfig}
\usepackage{amsmath}
\usepackage{graphicx}
\usepackage{amssymb}
\usepackage{bm}
\usepackage{dcolumn}

\large
\newcommand{\beq}{\begin{equation}}
\newcommand{\eeq}{\end{equation}}
\newcommand{\ba}{\begin{array}}
\newcommand{\ea}{\end{array}}
\newcommand{\bea}{\begin{eqnarray}}
\newcommand{\eea}{\end{eqnarray}}
\def\nn{\nonumber}
\def\bd{B_d^0}
\def\bs{B_s^0}
\def\bdbar{{\bar B}_d^0}
\def\bsbar{{\bar B}_s^0}
\def\ctl       {\ensuremath{\cos{\theta_l}}}
\def\stl       {\ensuremath{\sin{\theta_l}}}
\def\ctk       {\ensuremath{\cos{\theta_K}}}
\def\sstk       {\ensuremath{\sin^2{\theta_K}}}
\def\cstk       {\ensuremath{\cos^2{\theta_K}}}
\def\sstl       {\ensuremath{\cos2{\theta_l}}}
\def\sstl       {\ensuremath{\sin^2{\theta_l}}}

\def\cttl       {\ensuremath{\cos2{\theta_l}}}
\def\sttl       {\ensuremath{\sin2{\theta_l}}}

\def\sttk       {\ensuremath{\sin2{\theta_K}}}
\def\cp       {\ensuremath{\cos{\phi}}}
\def\sp       {\ensuremath{\sin{\phi}}}
\def\stp       {\ensuremath{\sin2{\phi}}}
\def\ctp       {\ensuremath{\cos2{\phi}}}
\def \cot {{c_{12}}}      
\def \coh {{c_{13}}}      
\def \cof {{c_{14}}}      
\def \cth {{c_{23}}}      
\def \ctf {{c_{24}}}      
\def \chf {{c_{34}}}      
\def \sot {{s_{12}}}      
\def \soh {{s_{13}}}      
\def \sof {{s_{14}}}      
\def \sth {{s_{23}}}      
\def \stf {{s_{24}}}      
\def \shf {{s_{34}}}      
\def \doh {{\delta_{13}}}      
\def \dof {{\delta_{14}}}      
\def \dtf {{\delta_{24}}}

\def\gsim{{~\raise.15em\hbox{$>$}\kern-.85em
          \lower.35em\hbox{$\sim$}~}}
\def\lsim{{~\raise.15em\hbox{$<$}\kern-.85em
          \lower.35em\hbox{$\sim$}~}}

\newcommand{\gev}{\ensuremath{\mathrm{\,Ge\kern -0.1em V}}}

\begin{document} 

\title{Flavor signatures of isosinglet vector-like down quark model}

\author{Ashutosh Kumar Alok}
\email{akalok@iitj.ac.in}
\affiliation{Indian Institute of Technology Jodhpur, Jodhpur 342011, India}

\author{Subhashish Banerjee}
\email{subhashish@iitj.ac.in}
\affiliation{Indian Institute of Technology Jodhpur, Jodhpur 342011, India}

\author{Dinesh Kumar}
\email{dinesh09@iitb.ac.in}
\affiliation{Indian Institute of Technology Bombay, Mumbai 400076, India\\
Department of Physics, University of Rajasthan, Jaipur 302004, India}

\author{S. Uma Sankar}
\email{uma@iitb.ac.in}
\affiliation{Indian Institute of Technology Bombay, Mumbai 400076, India}

\date{\today} 
\preprint{}

\begin{abstract}
We consider a model where the standard model is extended by the
addition of a vector-like isosinglet down-type quark $b'$. 
We perform a $\chi^2$ fit to the flavor physics data and obtain 
the preferred central values along with errors of all the 
elements of the measurable $3 \times 4$ quark mixing matrix.  
The fit indicates that all the new-physics parameters are consistent with 
zero and the mixing of the $b'$ quark with the other three is constrained to be 
small. The current flavor physics data rules out possibility of detectable new physics signals
in most of the flavor physics observables. 
We also investigate possible deviations in the standard model $Wtb$ couplings and bottom quark coupling to 
Higgs boson. We find that these deviations are less than a percent level which is too small to be observed at the LHC with current precision.
\end{abstract}

\maketitle 

\newpage

\section{Introduction}
The standard model (SM) consists of three generations of quarks, with two quarks in each generation. However, there is no {\it a priori} reason for the number of quarks to be restricted to six. It may be possible to have heavier quarks whose effects have not been detected yet. The minimal extension of the SM in this direction can be obtained by adding a vector-like isosinglet quark, either up-type or down-type, to the SM particle spectrum 
\cite{vqm,Barenboim:1997pf,Barenboim:2000zz,Barenboim:2001,Hawkins:2002qb,AguilarSaavedra:2002kr,Alok:2010ij,Cacciapaglia:2011fx,Botella:2012ju,Okada:2012gy,Alok:2012xm,Aguilar-Saavedra:2013qpa,Ellis:2014dza,Alok:2015iha,Ishiwata:2015cga}. Such exotic fermions can appear in $E_6$ grand unified theories as well in models with large extra dimensions. Since these quarks are vector-like, they do not lead to chiral anomalies. Here we consider the extension of SM by adding an isosinglet vector-like down-type quark $b'$.

As of now there are no direct evidences of exotic quarks. The additional chiral quarks, such as perturbative SM with fourth generation is excluded at the level of 5$\sigma$ by the recent LHC data on Higgs searches, when combined with electroweak precision data and direct searches at the LHC \cite{Eberhardt:2012gv}. As vector like fermions do not receive their mass from a Higgs doublet, they are still allowed by the existing experimental data and hence keeps us interested. 

The ordinary quarks with charge (-1/3) mix with the $b'$.  Because the $b'_L$ has a different $I_{3L}$ from $d_L$, $s_L$ and $b_L$, $Z$-mediated flavor changing neutral current (ZFCNC) appear at the tree level in the left-handed sector.  Thus the quark level transitions such as $b\to s$, $b\to d$, $s \to d$ can occur at the tree level. The addition of a $b'$ quark to the SM leads to a quark mixing matrix which is the 3 $\times$ 4 upper submatrix of a 
4 $\times$ 4 quark-mixing matrix CKM4, which is an extension of the Cabibbo-Kobayashi-Maskawa (CKM) quark-mixing matrix in the SM. This model thus provides a self-consistent framework to study deviations of 3 $\times$ 3 unitarity of the CKM matrix as well as flavor changing neutral currents at tree level. 

Not all the elements of the CKM matrix are measured directly; the values of the elements $V_{tq}$ ($q = d,s,b$) are determined from decays involving loops and by using the unitarity of the $3 \times 3$ CKM matrix. Hence one expects that due to the non unitarity of the quark mixing matrix in the ZFCNC model, sizable departures from the SM predictions may be possible. In this paper, we explore the possibility of such deviations by performing a fit to 
current flavor physics data. 

The addition of isosinglet down-type quark $b'$ modifies the couplings of SM bottom quark with $W$, $Z$ and Higgs boson. The deviations, if measured, can provide indirect evidence of vector quarks. In this work we study such possible deviations and provide an upper bound on them. 

 The quark mixing matrix in the SM, which is 3 $\times$ 3 unitary matrix, 
is parametrized by three angles, $\theta_{12}$, $\theta_{13}$, and $\theta_{23}$ and the $CP$-violating phase
$\delta_{13}$. The parametrization of 4 $\times$ 4  unitary quark-mixing matrix requires three additional angles
$\theta_{14}$, $\theta_{24}$, and $\theta_{34}$ and two additional $CP$-violating phases $\delta_{14}$ and $\delta_{24}$. 
In our analysis we use an exact parametrization of the CKM4 matrix \cite{Botella:1985gb,Fritzsch:1986gv,Harari:1986xf}:
{\footnotesize      
\begin{equation}     
\hspace{-1cm}      
 V_{CKM4} = \left( \begin{array}{cccc}      
\cot \coh \cof  & \coh \cof \sot & \cof \soh e^{-i\doh} & \sof e^{-i\dof} \\      
& & &\\      
-\cth \ctf \sot -\cot \ctf \soh \sth e^{i\doh} & \cot \cth \ctf -\ctf \sot \soh \sth e^{i\doh} &  \coh \ctf \sth & \cof \stf e^{-i\dtf} \\      
-\cot \coh \sof \stf e^{i(\dof-\dtf)} & -\coh \sot \sof \stf e^{i(\dof-\dtf)} &      -\soh \sof \stf e^{-i(\doh+\dtf-\dof)} &  \\      
& & & \\      
-\cot \cth \chf\soh e^{i\doh} +\chf \sot \sth & -\cot \chf \sth - \cth \chf \sot \soh e^{i\doh} &  \coh\cth\chf & \cof \ctf \shf  \\      
-\cot \coh \ctf \sof \shf e^{i \dof} & -\cot \cth \stf\shf e^{i\dtf} & -\coh\sth\stf\shf e^{i\dtf} &      \\      
+\cth\sot\stf\shf e^{i\dtf}& -\coh \ctf \sot\sof\shf e^{i\dof} & -\ctf\soh\sof\shf e^{i(\dof-\doh)} &      \\      
+\cot \soh \sth\stf\shf e^{i(\doh+\dtf)} & +\sot \soh \sth \stf \shf e^{i(\doh+\dtf)} & & \\      
& & &\\      
-\cot \coh \ctf \chf \sof e^{i\dof } & -\cot \cth \chf \stf e^{i\dtf}+\cot \sth \shf &  -\coh \cth \shf & \cof \ctf \chf\\      
+\cot \cth \soh \shf e^{i \doh} & -\coh \ctf \chf \sot \sof e^{i \dof}& -\coh \chf \sth \stf e^{i\dtf} & \\      
+\cth \ctf \sot \stf e^{i\dtf}-\sot \sth \shf & +\cth \sot \soh \shf e^{i \doh} &  -\ctf \chf \soh \sof e^{i(\dof-\doh)} & \\      
+\cot \chf \soh \sth \stf e^{i(\doh+\dtf)} & +\chf \sot  \soh \sth \stf e^{i(\doh + \dtf)} & &      
\end{array} \right) 
\label{CKM4}     
\end{equation}      }
with $s_{ij}=\sin \theta_{ij}$ and $c_{ij}=\cos \theta_{ij}$. Thus all the elements of the measurable $3 \times 4$ sub-matrix of CKM4 are expressed in terms of the nine CKM4 parameters. All the flavor observables, in turn, can be written in terms of these measurable CKM4 elements.

In this work, we consider the following flavor observables:
\begin{enumerate}
\item
The six directly measured magnitudes of the CKM matrix elements,
\item indirect and direct $CP$ violation in $K_L \to \pi \pi$,
\item the branching ratio of $K^+ \to \pi^+ \nu \bar{\nu}$ and $K_L \to \mu^+  \mu^-$, 
\item various onservables in $Z \to b \bar{b}$ decay,
\item  $\bs$-$\bsbar$ and $\bd$-$\bdbar$ mixing,
\item 
the time-dependent indirect $CP$ asymmetries in $\bd \to J/\psi\, K_S$ and $\bs \to J/\psi\, \phi$,
\item
the measurement of the angle $\gamma$ of the unitarity triangle from {\em tree-level decays}, 
\item
the branching ratio of the inclusive decay $B \to X_s l^+ l^-$ and of the exclusive decay $B \to K
\mu^+ \mu^-$,
\item many observables in $B \to K^* \mu^+ \mu^-$,
\item the branching ratio of  $B^+ \to \pi^+ \mu^+ \mu^-$, 
\item the branching ratios of $\bs \to \mu^+ \mu^-$, $\bd \to \mu^+ \mu^-$ and $B^+ \to \tau^+ \nu_{\tau}$, 
\item the like-sign dimuon charge asymmetry $A^b_{SL}$,
\item the oblique parameters $S$, $U$ and $T$,  and
\item $D$-$\bar{D}$ mixing.
\end{enumerate}

We compare the measured values of the above quantities to the theoretical expressions for them in the standard CKM and do a $\chi^2$ fit to obtain the SM CKM parameters. Then we redo the fit, using the corresponding theoretical expressions in the isosinglet vector-like down-type quark model and obtain values for the SM CKM  parameters as well as the new physics magnitudes $\theta_{14},\, \theta_{24}$ and $\theta_{34}$ and the new physics phases $\delta_{14}$ and $\delta_{24}$.

We then turn on to predict observables that are expected to be affected by the $b'$ quark, 
while still being consistent with the above measurements. We examine following observables: 
(i) the branching fraction of $K_L \to \pi^0 \nu \bar{\nu}$,
(ii) the branching fraction of $B \to X_s \nu \bar{\nu}$,
(iii) direct $CP$ asymmetry in $B \to (K,\,K^*)\,\mu^+\,\mu^-$, and
(iv) deviations in the standard model $Wtb$ couplings
and bottom quark coupling to Higgs boson.

The paper is organized as follows.  In Sec.~\ref{zfcnc}, we define the model, list the input values of various quantities used in the fit and discuss the definitions of $\chi^2$ for each individual observable. The results of the fit are presented in Sec.~\ref{results}. Using the results of the fit, the predictions for several observables, which are to be measured, are given in Sec.~\ref{pred}. We conclude in Sec.~\ref{concl} with a discussion of the results. 
\section{Flavor changing couplings of Z boson to down-type quarks}
\label{zfcnc}
In SM the quark content is represented by:
\beq
 \begin{pmatrix}
u_{L} \\
d_{L} \\
\end{pmatrix},
u_{R},d_{R};
 \begin{pmatrix}
c_{L} \\
s_{L} \\
\end{pmatrix},
c_{R},s_{R};
 \begin{pmatrix}
t_{L} \\
b_{L} \\
\end{pmatrix},
t_{R},b_{R}.
\eeq
The left handed quarks are represented as doublets and the right handed quarks are represented as singlets under $SU(2)_{L}$. Here we extend the quark sector by adding an $SU(2)$ singlet vector-like quark of charge (-1/3), labelled $b'$. The mixing of this quark with the SM quarks of charge (-1/3) leads to a different structure for the CKM matrix. 
The $3 \times 3$ mixing matrix connecting the charge (2/3) quarks to the charge (-1/3) quarks of the SM is 
no longer unitary, but is a submatrix of a $4 \times 4$ unitary matrix.
Without loss of generality, we can choose the interaction and mass eigenbases of charge (2/3) quarks to be the same. 
Hence the up-type mass matrix is diagonal and real. The mass matrix of the charge (-1/3) quarks, in the interaction eigenbasis, is a general $4\times 4$ complex matrix $M$, which is put in a diagonal form by a bi-unitary transformation of the form $M_{dia} = V_L^\dagger M V_R$.  The unitary matrix $V_L$ appears in the charged current interactions, when they are rewritten in the quark mass eigenbases.  The first three rows of $V_L \equiv V$ are measureable in principle and the top $3 \times 3$ sub-block is no longer unitary. 
This leads the flavor changing couplings of the Z boson to the down-type quarks, which are given by
\beq
{\cal L}^Z_{FCNC} = -\frac{g}{2\cos\theta_W}U_{jk}\bar{d}_{jL}\gamma^\mu d_{kL}Z_\mu.
\eeq
$U_{jk}$ are defined in terms of the first three elements of the fourth row of $V_L$ as
$U_{ds} = -V^*_{4d}V_{4s},   U_{sb} = -V^*_{4s}V_{4b}$ and  $U_{db} = -V^*_{4d}V_{4b}$.

The current experimental values for the 72 flavor physics observables enumerated in the introduction  
are listed in Table \ref{tab1} and \ref{bkstar}. The theoretical expressions for these observables require additional inputs in the form of decay constants, bag parameters , QCD corrections and other parameters. These are listed in Table \ref{tab2}.
\begin{table}
\begin{center}
\begin{tabular}{|c|c|}
\hline
$|V_{ud}| = 0.97425\pm 0.00022$ & ${\cal{B}}(B\to X_s \ell^+ \ell^-)_{\rm low} = (1.60 \pm 0.48)\times 10^{-6}$\cite{Lees:2013nxa}\\
$|V_{us}| = 0.2252\pm 0.0009$ & ${\cal{B}}(B\to X_s \ell^+ \ell^-)_{\rm high} = (0.57 \pm 0.16)\times 10^{-6}$\cite{Lees:2013nxa}\\
$|V_{cd}| = 0.230\pm 0.011$ &$10^{9}\, {\rm GeV^2} \times \langle  \frac{d{\cal{B}}}{dq^2} \rangle(B\to K \mu^+ \mu^-)_{\rm low} = 18.7 \pm 3.6$\cite{Aaij:2014pli}\\
$|V_{cs}| = 1.006\pm 0.023$ &$10^{9}\, {\rm GeV^2} \times \langle  \frac{d{\cal{B}}}{dq^2} \rangle(B\to K \mu^+ \mu^-)_{\rm high} = 9.5 \pm 1.7$\cite{Aaij:2014pli}\\
$|V_{ub}| = 0.00382\pm 0.00021$ & ${\cal{B}}(B^+\to \pi^+ \mu^+ \mu^-) = (2.60 \pm 0.61)\times 10^{-8}$ \cite{LHCb:2012de}\\
$|V_{cb}| = (40.9\pm 1.0)\times 10^{-3}$  & ${\cal{B}}(K^+\to \pi^+\nu\bar\nu) = (1.7 \pm 1.1)\times 10^{-10}$ \\
$\gamma=(68.0 \pm 11.0)^{\circ}$ &  ${\cal{B}}(K_L\to \mu^+ \mu^-) \leq 2.5 \times 10^{-9}$ \cite{Isidori:2003ts}\\
$|\epsilon_K|\times 10^{3} = 2.228 \pm 0.011$ & ${\cal{B}}(B_s\to \mu^+ \mu^-) = (2.9\pm0.7)\times 10^{-9}$ \cite{Aaij:2013aka,Chatrchyan:2013bka,CMS:2014xfa} \\
$\epsilon'/\epsilon = (16.6 \pm 2.3) \times 10^{-4}$ & ${\cal{B}}(B_d\to \mu^+ \mu^-) = (3.9\pm1.6)\times 10^{-10}$ \cite{Aaij:2013aka,Chatrchyan:2013bka,CMS:2014xfa}\\
$\Delta{M_d} = (0.507 \pm 0.004)\, {\rm ps}^{-1}$\cite{Amhis:2012bh} &   ${\cal{B}}(B\to \tau \,\bar{\nu}) = (1.14 \pm 0.22)\times 10^{-4}$ \cite{Amhis:2012bh}\\
$\Delta{M_s} = (17.72 \pm 0.04)\, {\rm ps}^{-1}$\cite{Amhis:2012bh}&$A^b_{sl}=(-4.96 \pm 1.69)\times 10^{-3}$ \cite{Abazov:2013uma}\\
$S_{J/\psi\,\phi}= 0.00 \pm 0.07$\cite{Amhis:2012bh}& $ S = 0.05 \pm 0.10$ \\
$S_{J/\psi\,K_S}= 0.68 \pm 0.02$\cite{Amhis:2012bh}& $ U = -0.03 \pm 0.10$ \\
$R_b = 0.21629 \pm 0.00066$\cite{ALEPH:2005ab} & $ T = 0.01 \pm 0.12$\\
$A_b^{FB} = 0.0992 \pm 0.0016$\cite{ALEPH:2005ab}  &  \\
$A_b = 0.923 \pm 0.020$\cite{ALEPH:2005ab}  &  \\
$R_c = 0.1721 \pm 0.003$\cite{ALEPH:2005ab}  & \\
\hline 
\end{tabular} 
\caption{Experimental values of flavor-physics observables used as
  constraints.  For $V_{ub}$ we use the weighted average from the
  inclusive and exclusive semileptonic decays, $V_{ub}^{inc}=(44.1\pm
  3.1) \times 10^{-4}$ and $V_{ub}^{exc}=(32.3 \pm 3.1)\times
  10^{-4}$. When not explicitly stated, the inputs are taken from the
  Particle Data Group \cite{pdg}. The asymmetric
  experimental errors are symmetrized by taking the largest side
  error. Also, wherever there is more than one source of uncertainty,
  the total error is obtained by adding them in quadrature.}
\label{tab1}
\end{center}
\end{table}

\begin{table}
\begin{center}
\begin{tabular}{|c|c|c|}
\hline
$q^2 = 0.1$-2 GeV$^2$ & $q^2 = 2$-4.3 GeV$^2$ & $q^2 = 4.3$-8.68 GeV$^2$  \\
\hline 
 $\langle\frac{d{\cal{B}}}{dq^2}\rangle$ = $(0.60 \pm 0.10)\times 10^{-7}$
& $\langle\frac{d{\cal{B}}}{dq^2}\rangle$ = $(0.30 \pm 0.05)\times 10^{-7}$ &$\langle\frac{d{\cal{B}}}{dq^2}\rangle$ = $(0.49 \pm 0.08)\times 10^{-7}$\\
$\langle F_{L} \rangle$ = $ 0.37 \pm 0.11$& $\langle F_{L} \rangle$ = $ 0.74 \pm 0.10$ & 
$\langle F_{L} \rangle$ = $ 0.57 \pm 0.08 $ \\
$\langle P_1 \rangle$ = $ -0.19 \pm 0.40 $ & $\langle P_1 \rangle$ = $ -0.29 \pm 0.65$  & 
$\langle P_1 \rangle$ = $ 0.36 \pm 0.31$ \\
$\langle P_2 \rangle$ = $ 0.03 \pm 0.15$ & $\langle P_2 \rangle$ = $ 0.50 \pm 0.08$ & 
$\langle P_2 \rangle$ = $ -0.25 \pm 0.08$ \\
$\langle P_4' \rangle$ = $0.00 \pm 0.52$ & $\langle P_4' \rangle$ = $0.74 \pm 0.60$ & 
$\langle P_4' \rangle$ = $1.18 \pm 0.32$ \\ 
$\langle P_5' \rangle$ = $0.45 \pm 0.24$ & $\langle P_5' \rangle$ = $0.29 \pm 0.40$ &
$\langle P_5' \rangle$ = $-0.19 \pm 0.16$ \\
$\langle P_6' \rangle$ = $0.24 \pm 0.23$ & $\langle P_6' \rangle$ = $-0.15 \pm 0.38 $ & 
$\langle P_6' \rangle$ = $0.04 \pm 0.16$ \\
$\langle P_8' \rangle$ = $-0.12 \pm 0.56$ & $\langle P_8' \rangle$ = $-0.3 \pm 0.60$ &
$\langle P_8' \rangle$ = $0.58 \pm 0.38$ \\
\hline
$q^2 = 14.18$-16 GeV$^2$ & $q^2 = 16$-19 GeV$^2$&  \\
\hline
 $\langle\frac{d{\cal{B}}}{dq^2}\rangle$ = $(0.56 \pm 0.10)\times 10^{-7}$
& $\langle\frac{d{\cal{B}}}{dq^2}\rangle$ = $(0.41 \pm 0.07)\times 10^{-7}$ &\\
 $\langle F_{L} \rangle$ = $ 0.33 \pm 0.09$
& $ \langle F_{L} \rangle$ = $ 0.38 \pm 0.09$ &\\
 $\langle P_1 \rangle$ = $ 0.07 \pm 0.28$
& $ \langle P_1 \rangle$ = $ -0.71 \pm 0.36$ &\\
 $\langle P_2 \rangle$ = $ -0.50 \pm 0.03$
& $ \langle P_2 \rangle$ = $ -0.32 \pm 0.08$ &\\
$\langle P_4' \rangle$ = $-0.18 \pm 0.70$
& $ \langle P_4' \rangle$ = $0.70 \pm 0.52$ &\\
 $\langle P_5' \rangle$ = $-0.79 \pm 0.27$
& $ \langle P_5' \rangle$ = $-0.60 \pm 0.21 $ &\\
$\langle P_6' \rangle$ = $0.18 \pm 0.25$
& $ \langle P_6' \rangle$ = $-0.31 \pm 0.39$ &\\
 $\langle P_8' \rangle$ = $-0.40 \pm 0.60$ 
& $ \langle P_8' \rangle$ = $0.12 \pm 0.54$ &\\
 \hline
\end{tabular}
\caption{ Experimental values of  $B \to K^*\, \mu^+\,\mu^-$ observables used as constraints. They are taken from
  Refs.~\cite{Aaij:2013iag,Aaij:2013qta}. Here the errors have been
  symmetrized by taking the largest side error. Also, wherever there
  is more than one source of uncertainty, the total error is obtained
  by adding them in quadrature.}
\label{bkstar}
\end{center}
\end{table}

\begin{table}
\begin{center}
\begin{tabular}{|c|c|}
\hline
$G_F = 1.16637 \times 10^{-5}$ Gev$^{-2}$&  $\tau_{K_L} = (5.116 \pm 0.021)\times 10^{-8}$ s \\
$ \sin^2\theta_W = 0.23116$              &  $\tau_{K^+} = (1.2380 \pm 0.0020)\times 10^{-8}$ s \\
$ \alpha(M_Z) = \frac{1}{127.9}$         &  $ \eta_c = 1.43\pm 0.23$ \cite{Herrlich:1996vf}  \\
$ \alpha_s(M_Z) = 0.1184$                &  $ \eta_{ct} = 0.496 \pm 0.047$ \cite{Brod:2010mj}\\
$m_t(m_t) = 163 $ GeV                    &  $ \eta_t = 0.5765$ \cite{Buras:1990fn}  \\
$m_c(m_c) = 1.275 \pm 0.025 $ GeV        &  $f_K = 0.1561 \pm 0.0011$ \cite{Laiho:2009eu}\\
$m_b(m_b) = 4.18 \pm 0.03 $ GeV          &  $\hat{B}_K = 0.767 \pm 0.010$ \cite{Laiho:2009eu}\\
$M_W = 80.385$ GeV                       &  $\Delta M_K = (0.5292 \pm 0.0009)\times 10^{-2} \, {\rm ps}^{-1}$\\
$M_Z = 91.1876$ GeV                      &  $f_D = (0.209\pm0.003)$ GeV \cite{Aoki:2013ldr}\\
$M_K = 0.497614$ GeV                     &  $\hat{B}_D = 1.18\pm 0.07$ \cite{Buras:2010nd}\\
$M_{K^*} = 0.89594$ GeV                  &  $ \kappa_{\epsilon}=0.94 \pm 0.02$ \cite{Buras:2008nn,Buras:2010pza} \\
$M_{B_d} = 5.27917$ GeV                  &  $f_{bd}=(190.5 \pm 4.2)$ MeV \cite{Aoki:2013ldr}\\
$M_{B_s} = 5.36677$ GeV                  &  $f_{bs}=(227.7\pm4.5)$ MeV \cite{Aoki:2013ldr}\\
$M_{B^{\pm}} = 5.27926$ GeV              &  $f_{\bd}\sqrt{B_{\bd}} = (0.216 \pm 0.015)$ {\rm GeV}\cite{Aoki:2013ldr} \\
$M_D = 1.864$ GeV                        &  $f_{\bs}\sqrt{B_{\bs}}=(0.266 \pm 0.018)$ {\rm GeV} \cite{Aoki:2013ldr}\\
$m_{\mu} = 0.105$ GeV                    &  ${\cal{B}}(B\to X_c \ell \nu) = (10.61 \pm 0.17)\times 10^{-2}$\\
$m_{\tau} = 1.77682$ GeV                 &  ${\cal{B}}(K^+\to \pi^0 e^+ \nu) = ( 5.07\pm 0.04)\%$ \\
$\tau_{B_d}= (1.519\pm 0.007)$ ps        &  ${\cal{B}}(K^+\to \mu^+ \nu) = (63.56 \pm 0.11)\%$\\
$\tau_{B_s}=(1.497\pm 0.026)$ ps         &  $m_c/m_b=0.29 \pm 0.02$\\
$\tau_{B^{\pm}}=(1.641\pm0.008)$ ps      &  $\eta^Z_B=0.57 $ \cite{AguilarSaavedra:2002kr} \\
\hline
\end{tabular}
\caption{Decay constants, bag parameters, QCD corrections and other
  parameters used in our analysis.  When not explicitly stated, we take the
  inputs from the Particle Data Group \cite{pdg}.}
\label{tab2}
\end{center}
\end{table}

For the fit, we define the total $\chi^2$ function as 
\bea
\chi^2_{\rm total} & = & 
\chi^2_{\rm CKM} 
+ \chi^2_{|\epsilon_K|} 
+ \chi^2_{\epsilon'/\epsilon}
+ \chi^2_{K \to \pi^+ \nu \bar{\nu}} 
+ \chi^2_{K_L \to \mu^+ \mu^-}
+ \chi^2_{Z\to b{\bar b}}
+ \chi^2_{\bd} 
+ \chi^2_{\bs}
\nonumber\\
&& \hskip2truemm  
+~ \chi^2_{\sin 2\beta} 
+\chi^2_{\sin 2\beta_s} 
+ \chi^2_{\gamma} 
+ \chi^2_{B \to X_s\, l^+ \,l^-} 
+ \chi^2_{B \to K\, \mu^+ \,\mu^-} 
+ \chi^2_{B  \to K^* \, \mu^+ \,\mu^-} 
\nonumber\\
&& \hskip2truemm 
+~\chi^2_{B^+ \to \pi^+\, \mu^+\, \mu^-}
+\chi^2_{B_q \to \mu^+ \mu^-}
+ \chi^2_{B \to \tau \,\nu}
+ \chi^2_{A^b_{SL}}
+ \chi^2_{\rm Oblique}
+ \chi^2_{ D}~.
\eea
In our analysis $\chi^2$ of an observable $A$ is defined as
\beq
\chi^2_A = \left( \frac{A - A_{exp}^c}{A_{exp}^{err}} \right)^2,
\eeq
where the measured value of $A$ is $(A_{exp}^c \pm A_{exp}^{err})$.
The individual components of the function $\chi^2_{\rm total}$, 
i.e the $\chi^2$ of different observables that we are using as inputs, 
are defined in the following subsections.

\subsection{\bf \boldmath Direct measurements of the CKM elements }

The contribution to the $\chi^2$ from the direct measurements of the magnitudes of the CKM elements is given by
\bea
\chi^2_{\rm CKM} &=& \Big( \frac{|V_{us}|-0.2252}{0.0009} \Big)^2 
+ \Big( \frac{|V_{ud}|-0.97425}{0.00022} \Big)^2
+\Big( \frac{|V_{cs}|-1.006}{0.023} \Big)^2 
\nonumber\\&&
+~\Big( \frac{|V_{cd}|-0.230}{0.011} \Big)^2
+\Big( \frac{|V_{ub}|-0.00382}{0.00021} \Big)^2 
+ \Big( \frac{|V_{cb}|-0.0409}{0.001} \Big)^2 \;.
\eea
%

\subsection{\bf \boldmath Indirect CP violation $\epsilon_K$ in $K_L \to \pi\pi$}

The mixing induced $CP$ asymmetry in neutral $K$ decays is described by the 
parameter $|\epsilon_K|$, which is proportional to ${\rm Im} (M^{12}_K)$. 
To calculate the contribution to $\chi^2$ from $|\epsilon_K|$,
we use the quantity 
\beq
K_{\rm mix} = \frac{12\sqrt{2}\pi^2(\Delta M_K)_{\rm exp} |\epsilon_K|}
{G^2_F M^2_W f^2_K m_K \hat{B}_K k_\epsilon}
\eeq
With the theoretical and experimental inputs given in Table~ \ref{tab1} and  \ref{tab2}, 
we find
\beq
K_{\rm mix,\, exp} = (1.69\pm0.05) \times 10^{-7} \; .
\eeq 
The contribution to $\chi^2$ from $|\epsilon_K|$ is then 
\beq
\chi^2_{\rm |\epsilon_K|} = \Big( \frac{K_{\rm mix} - 1.69 \times 10^{-7}}
{0.05 \times 10^{-7}} \Big)^2 + \chi^2_{\eta}\;,
\eeq
where  
\beq
\chi^2_{\eta} = \Big( \frac{\eta_{c} - 1.43}{0.23} \Big)^2 + \Big( \frac{\eta_{ct} - 0.496}{0.047} \Big)^2\;.
\eeq
Using the expression for $|\epsilon_K|$ given in 
\cite{Hawkins:2002qb}, it is 
straightforward to find an expression for $K_{\rm mix}$. 
In order to take into account the error in the QCD corrections $\eta_{c}$ and $\eta_{ct}$ which appear in the theoretical expression of $|\epsilon_K|$, we consider them to be parameters and have added a term, $\chi^2_{\eta}$, in $\chi^2$. We held the other QCD correction $\eta_t$ fixed to its central value because its error is very small.

\subsection{\bf \boldmath Direct CP violation $\epsilon'/\epsilon$ in $K_L \to \pi\pi$}

 The ratio $\epsilon'/\epsilon$ measures direct $CP$ violation in $K_L \to \pi\pi$
and has been measured quite accurately by NA48 \cite{Batley:2002gn} and KTeV \cite{AlaviHarati:2002ye,Worcester:2009qt} collaborations. The current
world average is $(16.6 \pm 2.3) \times 10^{-4}$. However, the SM prediction is subject to 
large uncertainties. Within the SM there is destructive interference between the QCD
penguins and the electroweak penguins  contributions. This one hand makes the theoretical predictions challenging
but on the other hand makes this observable sensitive to new physics which, in general, 
is expected to contribute to $Z$ penguins rather than the QCD penguins. Therefore in spite of
large theoretical uncertainties, $\epsilon'/\epsilon$ is expected to provide useful 
constraints on new physics parameters \cite{Buras:1998ed,Buras:2015yca}. 
This ratio is sensitive to $Im\,(U_{sd})$ \cite{Barenboim:2001,AguilarSaavedra:2002kr} and hence is
included in our analysis. 

 The dominant sources of uncertainties in the theoretical prediction of $\epsilon'/\epsilon$ 
is due to two non-perturbative parameters $B_6^{1/2}$  and $B_8^{3/2}$ that parametrise the matrix 
elements of the dominant operators $Q_6$ and $Q_8$, respectively. These parameters are
calculated within the framework of lattice QCD or the large $N$-approach \cite{Bardeen:1986uz,Buras:2014maa}. Using the 
recent results by the RBC-UKQCD lattice collaboration \cite{Blum:2015ywa,Bai:2015nea}, $(\epsilon'/\epsilon)_{\rm SM}$ is predicted 
to be  $(1.9 \pm 4.5) \times 10^{-4}$ \cite{Buras:2015yba} which is substantially more precise than the previous 
estimates of $(\epsilon'/\epsilon)_{\rm SM}$ and differs from the experimental measurement at
the level of 3$\sigma$. 

 The contribution to $\chi^2$ from $\epsilon'/\epsilon$ is given by
\begin{equation}
\chi^2_{\epsilon'/\epsilon} = \left(\frac{\epsilon'/\epsilon - 16.6 \times 10^{-4}}{2.3 \times 10^{-4}} \right)^2 + \chi^2_{th}\,,
\end{equation}
where 
\begin{eqnarray}
\chi^2_{th} &=& \left(\frac{B_6^{1/2} - 0.57}{0.19} \right)^2 + \left(\frac{B_8^{3/2} - 0.76}{0.05} \right)^2 \nonumber\\
&& + \left(\frac{\hat{\Omega}_{\rm eff} - 14.8 \times 10^{-2}}{8 \times 10^{-2}} \right)^2 + \left(\frac{a_0^{1/2} - (-2.92)}{0.12} \right)^2\,.
\end{eqnarray}
In order to include the error in quantities  $B_6^{1/2}$, $B_8^{3/2}$, $\hat{\Omega}_{\rm eff}$ and $a_0^{1/2}$ which appear in the theoretical expression of $\epsilon'/\epsilon$, the term $\chi^2_{th}$ is added to $\chi^2_{\epsilon'/\epsilon}$. The theoretical expression for $\epsilon'/\epsilon$ in ZFCNC model is taken from Ref.~\cite{Barenboim:2001,AguilarSaavedra:2002kr} whereas the numerical values of the theoretical inputs are taken from \cite{Buras:2015yba}.

\subsection{\bf \boldmath Branching fraction of the decay $K^+ \to \pi^+ \nu {\bar \nu}$}

Unlike other K decays, $K^+ \to \pi^+ \nu {\bar \nu}$ is dominated by the
short-distance (SD) interactions. The LD contribution to 
$K^+ \to \pi^+ \nu {\bar \nu}$ is about 3 orders of magnitude
smaller than that of the SD \cite{Rein:1989tr, Hagelin:1989wt}.

In order to include ${\cal B}(K^+ \to \pi^+ \nu {\bar \nu})$, we define
\beq
\chi^2_{K^+\to \pi^+\nu \bar{\nu}} =\Big(\frac{ K_{\rm slep} - 7.37\times 10^{-5}}{4.77\times 10^{-5}} \Big)^2 + \chi^2_{X}\;,
\eeq
where
\beq
\chi^2_{X} = \Big( \frac{X^{nl}_{e} - 10.6 \times 10^{-4}}
{1.5 \times 10^{-4}} \Big)^2 
+ \Big( \frac{X^{nl}_{\tau} - 7.1 \times 10^{-4}}
{1.4 \times 10^{-4}} \Big)^2\;,
\eeq

Using Table~\ref{tab1} and \ref{tab2}, we obtain
\beq
K_{\rm slep} = \frac{2\pi^{2} \sin^{4}\theta_W {\cal B}(K^+ \to \pi^+ \nu {\bar \nu})}{\alpha^2 r_K {\cal B}(K^+ \to \pi^0 e^+\nu )} = 
(7.37\pm 4.77)\times 10^{-5},
\eeq 
Here we have used $r_{K^+}=0.901\pm0.027$ which epitomizes the isospin-breaking corrections in
relating the branching ratio of $K^+\to \pi^+\nu\bar{\nu}$ to that of
the well-measured leading decay $K^+\to \pi^0 e^+ \nu$. Using the expression for ${\cal B}(K^+ \to \pi^+ \nu {\bar \nu})$ given in  \cite{Hawkins:2002qb}, it is straightforward to find an expression for $K_{\rm slep}$.  In order to include the error in quantities $X^{nl}_{e}$ and $X^{nl}_{\tau}$ which appear in the theorectical expression of  ${\cal B}(K^+ \to \pi^+ \nu {\bar \nu})$, we consider them to be  parameters and have added a term, $\chi^2_{X}$, in $\chi^2$.

\subsection{ \bf \boldmath Branching fraction of the decay $K_L \to \mu^+ \mu^-$}

Unlike $K^+ \to \pi^+ \nu {\bar \nu}$ , $K_L \to \mu^+ \mu^-$ is not dominated by clean SD
effects. The LD and SD contributions are comparable in size. In order to extract
bounds on the SD contribution to the branching ratio of $K_L \to \mu^+ \mu^-$ , it is extremely
important to have a theoretical control on the $K_L \to \gamma \gamma$ form factors with off-shell
photons. A conservative bound of $2.5 \times 10^{-9}$ on ${\cal B}(K_L \to \mu^+ \mu^-)$ from SD was obtained
in Ref. \cite{Isidori:2003ts}. We use this bound to constrain the ZFCNC parameters.
In order to include ${\cal B}(K_L \to \mu^+ \mu^-)$, we define
\beq
\chi^2_{K_L \to \mu^+ \mu^-} =\Big(\frac{ K_{\rm lep} - 3.39\times 10^{-6}}{3.78\times 10^{-6}} \Big)^2 + \chi^2_{Y_{NL}}\;,
\eeq
where
\beq
\chi^2_{Y_{NL}} =\Big(\frac{Y_{NL} - 2.94 \times 10^{-4}}
{0.28\times 10^{-4}} \Big)^2\;,
\eeq
Using the input Table~\ref{tab1}, we obtain
\beq
K_{\rm lep} = \frac{\pi^{2} \sin^{4}\theta_W {\cal B}(K_L \to \mu^+ \mu^-) \tau_{K^+}}{\alpha^2 {\cal B}(K^+ \to \mu^+ \nu )\tau_{K_L}} = 
(3.39\pm 3.78)\times 10^{-6}.
\eeq 
Using the expression for ${\cal B}(K_L \to \mu^+ \mu^-)$ given in  \cite{Hawkins:2002qb}, the theoretical expression for $K_{\rm lep}$ can be easily obtained. The quantity $Y_{NL}$ appears in the theoretical expression for ${\cal B}(K_L \to \mu^+ \mu^-)$.  In order to include error in $Y_{NL}$, we consider it to be a parameter and have added a term, $\chi^2_{Y_{NL}}$, in $\chi^2$.

\subsection{\bf \boldmath $Z \to b\bar{b}$ decay} 
The $b-b'$ mixing in ZFCNC model modifies the $Zb\bar{b}$ coupling at the tree level. This affects observables 
such as $R_b$, $A^b_{FB}$, $A_b$ and $R_c$.  The theoretical expressions of these observables in the ZFCNC model are given by \cite{Aguilar-Saavedra:2013qpa}
\begin{eqnarray}
R_b &=& R^{SM}_b \left(1-1.820\, |V_{4b}|^2\right),\nonumber\\
A^b_{FB} &=& A^{b,SM}_{FB} \left(1-0.164\, |V_{4b}|^2\right),\nonumber\\
A_b &=& A^{SM}_b \left(1-0.164\, |V_{4b}|^2\right),\nonumber\\
R_c &=& R^{SM}_c \left(1-0.500\, |V_{4b}|^2 \right),
\end{eqnarray}
where the SM predictions are obtained from a fit in Ref.~\cite{pdg}.
The $\chi^2$ contribution is then given by
\beq
\chi^2_{Zb\bar{b}} = \Big( \frac{R_{b}-0.21629}{0.00066} \Big)^2+\Big( \frac{A^b_{FB}-0.0992}{0.0016} \Big)^2
+\Big( \frac{A_{b}-0.923}{0.020} \Big)^2+\Big( \frac{R_{c}-0.1721}{0.003} \Big)^2\;.
\eeq

\subsection{\bf \boldmath $B^0_q$-$\bar B^0_q$ mixing ($q = d,s$)}

The theoretical expressions for $M^q_{12}$ ($q = d,s$) in the ZFCNC model is given by \cite{Barenboim:1997pf}
\beq
M^q_{12} = \frac{G^2_F M_W^2 M_{B_q} f_{bq}^2 \hat{B}_{bq}}{12 \pi^2}  \left[\left(V^*_{tq}V_{tb}\right)^2-a\left(V^*_{tq}V_{tb}\right)U_{qb}+b\, U^2_{qb} \right]\,,
\eeq
where
\beq
a=8\frac{Y(x_t)}{S(x_t)}, \qquad \qquad \qquad b=\frac{2\sqrt{2}\pi^2}{G_F M_W^2 S(x_t)}\frac{\eta^Z_B}{\eta_B}\,.
\eeq
Here $S(x_t)$ and $Y(x_t)$ are the Inami-Lim functions 
\cite{Inami:1980fz}, while $\eta_B$  and $\eta^Z_B$ are  the QCD correction factors.
To calculate $\chi^2_{B_q}$ for $B_q$-$\bar{B_q}$ mixing, we use the quantity
\beq
B^q_{\rm mix} = \frac {6\pi^2 \Delta M_q}{G^2_F M_W^2 M_{B_q} 
\hat{B}_{bq} f_{B_q}^2 \eta_B S(x_t)} \; .
\eeq
With the inputs given in Table~\ref{tab1}, we get 
\begin{align}
B^d_{\rm mix,{\rm exp}} &=  (6.56 \pm 0.77 ) \times 10^{-5},\\ 
B^s_{\rm mix,{\rm exp}} &=  (1.48 \pm 0.14 ) \times 10^{-3}. 
\end{align}
Then one gets
\begin{align}
\chi^2_{\bd} = \Big( \frac{B^d_{\rm mix}-6.56 \times 10^{-5}}{0.77 \times 10^{-5}} \Big)^2,\\
\chi^2_{\bs} = \Big( \frac{B^s_{\rm mix}-1.48 \times 10^{-3}}{0.14 \times 10^{-3}} \Big)^2.
\end{align}

\subsection{\bf \boldmath Indirect $CP$ violation in $ \bd \to J/\psi\, K_S$ and $ \bs \to J/\psi\, \phi$}
 In the SM, indirect CP violation in $ \bd \to J/\psi\, K_S$ and $ \bs \to J/\psi\, \phi$
probes $\sin 2\beta$ and $\sin 2\beta_s$, respectively. With NP, we
have
\beq
S_{J/\psi\, K_S} = \frac{{\rm Im}(M^d_{12})}{|M^d_{12}|}, \hskip 30pt
S_{J/\psi\, \phi} = -\frac{{\rm Im}(M^s_{12})}{|M^s_{12}|}\;.
\eeq
The theoretical expressions for $M^q_{12}$ ($q = d,s$) in the ZFCNC
model are given in the previous subsection.
Using the  experimentally-measured values of $S_{J/\psi\, K_S}$ and 
$S_{J/\psi\, \phi}$ given in Table~\ref{tab1}, we get
\beq
\chi^2_{\sin 2\beta} = \Big( \frac{S_{J/\psi\, K_S}-0.68}{0.02}\Big)^2, \hskip 30pt
\chi^2_{\sin 2\beta_s} = \Big( \frac{S_{J/\psi\, \phi}-0.00}{0.07} \Big)^2 ~.
\eeq
\subsection{ \bf \boldmath CKM angle $\gamma$}

In the Wolfenstein parametrization, the CKM angle $\gamma=\tan^{-1}(\eta / \rho)$, 
which is the argument of $V_{ub}$. Therefore
the $\chi^2$ of $\gamma$ is given by
\beq
\chi^2_{\gamma} = \Big( \frac{\delta_{13}- 68~(\pi/180)}{11~(\pi/180)} \Big)^2\;.
\eeq

\subsection{ \bf \boldmath Branching ratio of $ B \to X_s\, l^+ \,l^-$ }
The effective Hamiltonian for the quark-level transition
$b \to s\,  l^+ \,l^-$ in the SM can be written as
\beq
{\cal H}_{eff} =  - \frac{4 G_F}{\sqrt{2}} V^*_{ts}V_{tb}
\sum_{i=1}^{10} C_i(\mu) \,  O_i(\mu)\;,
\label{Heffbs}
\eeq
where the form of the operators $O_i$ and the expressions for
calculating the coefficients $C_i$ are given in
Ref.~\cite{Buras:1994dj}. The $Z{\bar b}s$ coupling generated in the $Z$FCNC model changes the values of the 
Wilson coefficients $C_{9,10}$.  The Wilson coefficients $C^{\rm tot}_{9,10}$ in
the $Z$FCNC model can be written as \cite{Alok:2012xm}
\bea
C^{\rm tot}_{9} &=& C_9^{\rm eff} - \frac{\pi}{\alpha}\frac{U_{sb}}{V^*_{ts}V_{tb}} (4\sin^2 \theta_{W}-1)\,\nn\\
C^{\rm tot}_{10} &=& C_{10}-\frac{\pi}{\alpha}\frac{U_{sb}}{V^*_{ts}V_{tb}}\,.
\label{ctot}
\eea

The theoretical prediction for the branching fraction of $ B \to X_s \mu^+ \,\mu^-$
in the intermediate $q^2$ region ($7$~GeV$^2 \le q^2 \le 12$~GeV$^2$) 
is rather uncertain due to the nearby charmed resonances. 
The predictions are relatively cleaner in the low-$q^2$ ($1 \,{\rm GeV^2} 
\le q^2 \le 6\, {\rm GeV^2}$) and the high-$q^2$
($14.2\, {\rm GeV^2} \le q^2 \le m_b^2$) regions. We therefore consider
both low-$q^2$ and high-$q^2$  regions in the fit. 
The latest Belle measurement uses only
~25\% of its final data set \cite{Iwasaki:2005sy}. The BaBar
Collaboration has recently updated the measurement of ${\cal B}(B
\to X_s \, l^+ \,l^-)$ using the full data set, which corresponds
to $471 \times 10^{6}$ $B\bar{B}$ events \cite{Lees:2013nxa}. 

The theoretical predictions for ${\cal B}(B \to X_s \, l^+ \,l^-)$ are computed using
the program {\bf SuperIso} \cite{Mahmoudi:2007vz,Mahmoudi:2008tp}, in
which the higher-order and power corrections are taken from 
Refs.~\cite{Ghinculov:2003qd, Huber:2005ig}, while the electromagnetic
logarithmically-enhanced corrections and Bremsstrahlung contributions are
implemented following Refs.~\cite{Huber:2007vv} and \cite{Asatryan:2002iy},
respectively.

The contribution to $\chi^2_{\rm total}$ is
\bea
\chi^2_{B \to X_s\, l^+ \,l^-} & = & \Big( \frac{{\cal B}(B  \to X_s \, l^+ \,l^-)_{\rm low}-1.6\times 10^{-6}}{0.49\times 10^{-6}} \Big)^2 \nonumber\\
&& 
+~\Big( \frac{{\cal B}(B  \to X_s \, l^+ \,l^-)_{\rm high}-0.57\times 10^{-6}}{0.23\times 10^{-6}} \Big)^2 ~,
\eea
where we have added a theoretical error of $7\%$ to ${\cal B}(B \to
X_s\,l^+ \,l^-)_{\rm low}$, which includes corrections due to the
renormalization scale and quark masses, and a theoretical error of
$30\%$ to ${\cal B}(B \to X_s\,l^+ \,l^-)_{\rm high}$, which includes
the non-perturbative QCD corrections.

\subsection{\bf \boldmath Branching ratio of $B \to K\, \mu^+ \,\mu^-$}

The predictions for the branching ratio of $B \to K\, \mu^+ \,\mu^-$ are relatively 
cleaner in the low-$q^2$ (1.1 $ \rm GeV^2$ $\leq$ $q^2$ $ \leq$ 6 $ \rm GeV^2$) 
and the high-$q^2$ (15 $ \rm GeV^2$ $ \leq$ $q^2$ $ \leq$ 22 $ \rm GeV^2$) regions. 
We include both regions in the fit. We use the recent LHCb measurements
of $\langle d{\cal B}/dq^2\rangle(B \to K \, \mu^+ \,\mu^-)$
\cite{Aaij:2014pli}. The theoretical expression for $\langle d{\cal B}/dq^2\rangle(B \to K
\, \mu^+ \,\mu^-)$ in the SM are taken from Refs.~\cite{Bobeth:2007dw,Bobeth:2011nj} 
modulo the modified Wilson coefficients given in Eq.~\ref{ctot}.  

We include factorizable and non-factorizable corrections of $O(\alpha_s)$ in our numerical analysis following
Refs.~\cite{Bobeth:2007dw,Beneke:2001at} in the low-$q^2$ region. 
In the high-$q^2$ region, we make use of the improved Isgur-Wise
relation between the form factors \cite{Bobeth:2011nj}.  
The contribution to $ \chi^2_{\rm total}$ from $B \to K \, \mu^+ \,\mu^-$ is
\bea
\chi^2_{B \to K\, \mu^+ \,\mu^-} & = & \Big( \frac{\langle \frac{d{\cal B}}{dq^2}\rangle(B \to K \, \mu^+ \,\mu^-)_{\rm low}-18.7\times 10^{-9}}{6.67\times 10^{-9}} \Big)^2 \nn\\
&& \hskip2truecm 
+~\Big( \frac{\langle \frac{d{\cal B}}{dq^2}\rangle(B \to K \, \mu^+ \,\mu^-)_{\rm high}-9.5\times 10^{-9}}{3.32\times 10^{-9}} \Big)^2 ~,
\eea
where we have included a theoretical error of $30\%$ in both low- and high-$q^2$
bins. This is mainly due to uncertainties in the $B \to K$ form
factors.

\subsection{\bf \boldmath Constraints from  $B \to K^*\, \mu^+ \,\mu^-$}
A possible indicator of new physics in $b \to s$ sector could be the 
measurement of new angular observables in $B \to K^*\, \mu^+ \,\mu^-$ at the LHCb \cite{Aaij:2013qta,Descotes-Genon:2013wba}.
Here, we include all measured observables in $B \to K^*\, \mu^+
\,\mu^-$ in the low- and high-$q^2$ regions. The experimental results
for $B \to K^*\, \mu^+ \,\mu^-$ decay are given in Table \ref{bkstar}.

The complete angular distribution for the decay $B \to K^*\, \mu^+
\,\mu^-$ is described by four independent kinematic variables: the
lepton-pair invariant mass squared $q^2$, two polar angles
$\theta_\mu$ and $\theta_K$, and the angle between the planes of the
dimuon and $K\pi$ decays, $\phi$. The differential decay distribution
of $B \to K^*\, \mu^+ \,\mu^-$ can be written as
\begin{equation}
  \label{eq:differential decay rate}
  \frac{d^4\Gamma[B \to K^{*}(\to K \pi)\mu^+\mu^-]}
       {d q^2\, d\ctl\, d\ctk\, d\phi} =
  \frac{9}{32\pi}  J(q^2 , \theta_l, \theta_K, \phi)\,.
\end{equation}
where the angular-dependent term can be written as 
\bea
 J(q^2 , \theta_l, \theta_K, \phi)&=&J_{1s}\sstk + J_{1c}\cstk + (J_{2s}\sstk + J_{2c}\cstk)\cttl \nonumber \\
&& \hskip-2truecm +~J_3\sstk\sstl\ctp \nonumber + J_4\sttk\sttl\cp \nonumber \\
&& \hskip-2truecm +~J_5\sttk\stl\cp + (J_{6s}\sstk + J_{6c}\cstk)\ctl \\
&& \hskip-2truecm +~J_7\sttk\stl\sp + J_8\sttk\sttl\sp + J_9\sstk\sstl\stp ~. \nn
\eea
The $J_i$'s depend on the six complex $K^*$ spin
amplitudes $A_{\parallel}^{L,R}, A_{\perp}^{L,R}$, $A_0^{L,R}$ and $A_t$. For
example,
\begin{equation}
J_{1s} = \frac{(2+\beta_l^2)}{4}[|A_{\perp}^{L}|^2+|A_{\parallel}^{L}|^2+|A_{\perp}^{R}|^2+|A_{\parallel}^{R}|^2] + 
\frac{4m_l^2}{q^2}Re(A_{\perp}^{L}A_{\perp}^{R*}+A_{\parallel}^{L}A_{\parallel}^{R*}) ~.
\end{equation}
We can also define the optimized observables like $P_1$, $P_2$, $P'_4$, $P'_5$, $P'_6$, $P'_8$ \cite{Descotes-Genon:2013vna}. 
These observables are form factor independent
observables and having reduced hadronic uncertanities at leading order in corresponding effective-theory expansions. These form factor independent observables integrated in $q^2$ bins can be defined as, for example:
$$<P_1>_{bin} = \frac{1}{2}\frac{\int\limits_{bin}dq^2[J_3+\bar{J_3}]}{\int\limits_{bin}dq^2[J_{2s}+\bar{J_{2s}]}}\, $$
where $ \bar{J}_{i}$'s can be obtained from $J_i$'s by all weak phases conjugated.

For $B \to K^* \, \mu^+ \,\mu^-$, we use the observables $\langle d{\cal B}/dq^2\rangle$, 
$P_1$, $P_2$, $P'_4$, $P'_5$, $P'_6$, $P'_8$ and $F_L$ in the low-$q^2$ bins 0.1-2 GeV$^2$,
2.0-4.3 GeV$^2$, 4.3-8.68 GeV$^2$, and the high-$q^2$
bins 14.18-16 GeV$^2$ and 16-19 GeV$^2$. The observables $A_{FB}, F_L $ and $P_2$ are related as
$A_{FB} = -\frac{3}{2}(1-F_L)P_2$. These observables are highly correlated in most of the bins \cite{Hurth:2013ssa}.
This is the reason why we use $F_L $, instead of $A_{FB}$, in the fit as it does not show a strong correlation with $P_2$.
 The SM theoretical expressions for all observables in $B \to K^*\, \mu^+ \,\mu^-$ 
 are given in \cite{Descotes-Genon:2013vna} and could be adapted to the ZFCNC model by modification of the 
Wilson coefficients values, Eq.~(\ref{ctot}). 
These  predictions have errors associated with them. Excluding 
uncertainties due to CKM matrix elements, the main sources of uncertainties in the low-$q^2$ region are the form factors,
 unknown $1/m_b$ subleading corrections, quark
masses, and the renormalization scale $ \mu_b$.  Also, in the
high-$q^2$ region, there is an additional subleading correction of
$O(1/m_b)$ to the improved Isgur-Wise form factor relations. The theoretical error
for each $B \to K^*\,\mu^+ \,\mu^-$ observable $O_j$,  is
incorporated in the fit by multiplying the theoretical result by
$(1\pm X_j)$, where $X_j$ is the total theoretical error corresponding
to the $j^{\rm th}$ observable. This can be easily estimated using Table
II of Ref.~\cite{Hurth:2014vma}.
The theoretical predictions for all $B \to K^* \, \mu^+ \,\mu^-$ observables
are computed using the program {\bf SuperIso}
\cite{Mahmoudi:2007vz,Mahmoudi:2008tp}.

For each bin, we compute the flavor observables.  The $\chi^2$, which includes the experimental correlations, is defined as
\begin{equation}
\chi^2_{B  \to K^* \, \mu^+ \,\mu^-}  =
\sum_{\rm bins} \quad \Bigl[\sum_{j,\,k \in ({B\to K^* \mu^+ \mu^- \,{\rm obs.}})}\Bigl(O_j^{\rm exp} - O_j^{\rm th}\Bigr)\Bigl(\sigma^{bin}\Bigr)^{-1}_{jk}\Bigl(O_k^{\rm exp} - O_k^{\rm th}\Bigr)\Bigr]\,,
\end{equation}
where $\Bigl(\sigma^{bin}\Bigr)^{-1}_{jk}$ are the inverse of the covariance matrices for each bin which are computed using the correlation matrices given in Ref.~\cite{Hurth:2013ssa}.

\subsection{\bf \boldmath Branching ratio of $B^+ \to \pi^+ \,\mu^+ \, \mu^-$ }
The decay $B^+ \to \pi^+ \,\mu^+ \, \mu^-$ is the first measurement of any decay channel 
induced by $b \to d \,\mu^+\, \mu^-$. The measured branching ratio of $B^+ \to \pi^+ \,\mu^+ \, \mu^-$ is
$(2.3 \pm 0.6 \pm 0.1) \times 10^{-8}$ \cite{LHCb:2012de}.
The effective Hamiltonian for the quark level transition $b \to d\, \mu^+\, \mu^-$
along with the modified Wilson coefficients in the ZFCNC model can be
respectively obtained from Eqs.~(\ref{Heffbs}) and (\ref{ctot}) by
replacing $s$ by $d$. The theoretical expression for 
${\cal B}(B^+ \to \pi^+\, \mu^+\, \mu^-)$ in the ZFCNC model is obtained using the expressions given in Ref.~\cite{Wang:2007sp}.
The contribution to $\chi^2_{\rm total}$ is
\beq
\chi^2_{B^+ \to \pi^+\, \mu^+\, \mu^-}  =\Big( \frac{{\cal B}(B^+ \to \pi^+\, \mu^+\, \mu^-) - 2.3\times 10^{-8}}
{0.66\times 10^{-8}} \Big)^2\; ,
\eeq
where we have included a theoretical error of $10\%$ in ${\cal B}(B^+ \to \pi^+\, \mu^+\,
\mu^-)$ which is mainly is due to uncertainties in the $B^+ \to \pi^+$ form factors \cite{Ball:2004ye}.

\subsection{\bf \boldmath Branching ratio of $B_q \to \mu^+ \,\mu^-$ $(q=s,d)$ }
The branching ratio of $B_q \to \mu^+ \,\mu^-$ in the ZFCNC model is
given by
\beq
{\cal B}(B_q \to \mu^+ \,\mu^-) = \frac{G^2_F \alpha^2 M_{B_q} m_\mu^2 f_{bq}^2 \tau_{B_q}}{16 \pi^3} 
|V^*_{tq}V_{tb}|^2  \sqrt{1 - 4 (m_\mu^2/M_{B_q}^2)} 
|C^{\rm tot, q}_{10}|^2 ~,
\eeq
where $C^{\rm tot, s}_{10}$ is defined in Eq.~(\ref{ctot}), and
$C^{\rm tot, d}_{10}$ is given by
\beq 
C^{\rm tot, d}_{10} = C_{10}-\frac{\pi}{\alpha}\frac{U_{db}}{V^*_{td}V_{tb}}\,.
\eeq 
In order to include ${\cal B}(B_q \to \mu^+ \,\mu^-)$ $(q=s,d)$ in the
fit, we define
\beq
B_{\rm lepq} = \frac{16 \pi^3 {\cal{B}}( B_q \to \mu^+ \,\mu^-)}{G^2_F \alpha^2 M_{B_q} m_\mu^2 f_{bq}^2 \tau_{B_q}
 \sqrt{1 - 4 (m_\mu^2/M_{B_q}^2)}} \,.
\eeq
Using the inputs given in Tables \ref{tab1} and \ref{tab2}, we obtain
$B_{\rm leps, {\rm exp}} = 0.025 \pm 0.006$ and $B_{\rm lepd, {\rm exp}} = 0.0048\pm 0.0020$.
The contribution to $\chi^2_{\rm total}$ from ${\cal B}(\bs \to \mu^+
\,\mu^-)$ and ${\cal B}(\bd \to \mu^+ \,\mu^-)$ is then given by
\beq
\chi^2_{B_q \to \mu^+ \mu^-} = \Big( \frac{B_{\rm leps} - 0.025}{0.006} \Big)^2 + \Big( \frac{B_{\rm lepd} - 0.0048}{0.0020} \Big)^2\;.
\eeq

\subsection{\bf \boldmath Branching ratio of $B\to \tau \,\bar{\nu}$} 
The branching ratio of $B\to \tau \,\bar{\nu}$ is given by
\beq
{\cal{B}}(B\to \tau \,\bar{\nu}) = \frac{G^2_F  M_{B} m_\tau^2}{8\pi} \left(1- \frac{m_\tau^2}{M^2_{B}}\right)^2 f_{bd}^2 |V_{ub}|^2 \tau_{B^{\pm}}.
\eeq
In order to include ${\cal B}(B\to \tau \,\bar{\nu})$ in the fit, we
define
\beq
B_{\rm Btau-nu} = \frac{8\pi {\cal{B}}(B\to \tau \,\bar{\nu})}{G^2_F  M_{B} m_\tau^2 f_{bd}^2 \tau_{B}
 (1 - m_\tau^2/M_{B}^2)^2}\,.
\eeq
Using the inputs given in Tables \ref{tab1} and \ref{tab2}, we obtain
$B_{\rm Btau-nu, {\rm exp}} = (1.779\pm 0.352)\times 10^{-5}$.
The contribution to $\chi^2_{\rm total}$  from ${\cal B}(B\to \tau \,\bar{\nu})$ is then given by
\beq
\chi^2_{B \to \tau \,\nu} = \Big( \frac{B_{\rm Btau-nu} - 1.779\times 10^{-5}}{0.352\times 10^{-5}} \Big)^2.
\eeq

\subsection{\bf \boldmath Like-sign dimuon charge asymmetry $A^b_{SL}$}
The CP-violating like-sign dimuon charge asymmetry in the $B$ system
is defined as
\bea
 A^b_{SL} \equiv \frac{N_b^{++}  - N_b^{--}}{N_b^{++}  + N_b^{--}} ~,
\eea 
where $N_b^{\pm\pm}$ is the number of events of $b {\bar b} \to
\mu^{\pm} \mu^{\pm} X$. This asymmetry can be written as a linear combination
of the asymmetry in $B_d$ and $B_s$ sector:
\beq
A^b_{SL} = c^d_{SL} A^d_{SL} + c^s_{SL} A^s_{SL} ~,
\eeq
where $A^q_{SL} = {\rm Im}\Big(\Gamma^{(q)}_{12}/M_{12}^{(q)}\Big)$
$(q=s,d)$, with $c^d_{SL}=0.594 \pm 0.022$ and $c^s_{SL}=0.406 \pm
0.022$. $A^b_{sl}$ has been measured by the D\O\ Collaboration. The measured
value is $(-4.96 \pm 1.53 \pm 0.72)\times 10^{-3}$
\cite{Abazov:2013uma} which deviates by 2.7$\sigma$ from the SM
prediction of $A^b_{SL}$ which is $(-2.44 \pm 0.42) \times 10^{-4}$.

The theoretical expression for $A^q_{SL}$ is given in Ref.~\cite{Botella:2014qya}. The contribution
to $\chi^2$ from $A^b_{SL}$ is given by
\beq
\chi^2_{A^b_{SL}} = \Big( \frac{A^b_{SL} - (-4.96\times 10^{-3})}{1.69\times 10^{-3}} \Big)^2 + \chi^2_{c},
\eeq
where
\bea
\chi^2_{c} &=&  \Big( \frac{c^d_{SL} - 0.594}{0.022} \Big)^2 +\Big( \frac{c^s_{SL} - 0.406}{0.022} \Big)^2 \nn\\
&& +~\Big( \frac{a - 10.5}{1.8} \Big)^2 + \Big( \frac{b - 0.2}{0.1} \Big)^2 
+ \Big( \frac{c - (-53.3)}{12} \Big)^2 ~.
\eea
The term $\chi^2_{c}$ is added to include errors in $c^d_{SL}$ and $c^s_{SL}$ 
as well as in quantities $a$, $b$ and $c$ which appear in the theoretical expressions for $A^q_{SL}$ \cite{Botella:2014qya}.

\subsection{\bf \boldmath The oblique parameters $S$, $U$ and $T$} 
The contribution to $\chi^2$ from oblique parameters is given by
\begin{equation}
\chi^2_{\rm Oblique} = \Bigl(\frac{S - 0.05}{0.10}\Bigr)^2+ \Bigl(\frac{U - (-0.03)}{0.10}\Bigr)^2+ \Bigl(\frac{T - 0.01}{0.12}\Bigr)^2.
\end{equation}
The theoretical expressions for $S$, $U$ and $T$ given in Ref.~\cite{Lavoura:1992np}. 

\subsection{\bf \boldmath $D$-$\bar{D}$ mixing} 

The fit is expected to have very weak dependence on $b'$ mass as the theoretical expressions 
for all the observables discussed in the above subsections, except the oblique parameters, 
are independent of the mass of $b'$ quark. 
 In order to include the dependence of $b'$ mass in the fit, one
should include constraints from $D$-$\bar{D}$ mixing \cite{Golowich:2007ka}, despite the fact the we do not
have a reliable estimate of the SM contribution to $D-\bar{D}$ mixing 
\cite{Donoghue:1985hh,Georgi:1992as,Ohl:1992sr,Golowich:1998pz,Bigi:2000wn,Falk:2001hx,Bianco:2003vb,Falk:2004wg,Golowich:2005pt,Petrov:2013usa}. The new physics
contribution to $M_{12}^{D}$ in ZFCNC model, which is due to box diagram involving heavy $b'$, can be
reliably estimated \cite{Branco:1995us,Golowich:2007ka}.

In order to include constraints from $D-\bar{D}$ mixing, we follow \cite{Buras:2010nd} and use a model independent bound 
on the new physics mixing amplitude,$M_{12}^{D,\,NP}$, obtained in \cite{Ciuchini:2007cw}. The contribution to $\chi^2$ from $D-\bar{D}$ mixing is given by
\beq
\chi^2_{D} = \Big( \frac{D_{\rm mix} - 2.68\times 10^{-6}}{3.35\times 10^{-6}} \Big)^2 ,
\eeq
where
\beq
D_{\rm mix}=\frac{12 \pi^2  |M_{12}^{D,\,NP}| }{G^2_F f^2_D \hat{B_D} M_D M^2_W }= (2.76 \pm 3.43)\times 10^{-6}.
\eeq

\section{Results of the fit}
\label{results}

The results of these fits are presented in Table~\ref{table:parameters}.  
The results of the fit for the SM are consistent with those obtained in
Ref.~\cite{pdg}.
The results for ZFCNC model correspond to a $b'$ mass of 800 GeV and 1200 GeV. 
The best fit values of the parameters of the upper $3 \times 3$ sub-block of 
CKM4 matrix are not affected much by
the addition of a vector-like isosinglet down-type quark $b'$
and are essentially the same as the SM CKM fit parameters. 
On the other hand, the new real parameters $\theta_{14}$, $\theta_{24}$, $\theta_{34}$ are consistent with zero.  This also is consistent with the observation that no meaningful constraints are obtained on the new phases $\delta_{14}$ and $\delta_{24}$: since vanishing $\theta_{14}$, $\theta_{24}$ imply vanishing $V_{ub'}, V_{cb'}$, respectively, the phases of these two elements have no significance. Therefore we see that even if we invoke violation of unitarity by adding a vector isosinglet down-type quark $b'$ to the SM particle spectrum, the constraints coming from the flavor physics sector does not allow any sizable deviations from the unitarity of 3 $\times$ 3 CKM matrix.

\begin{table}
\begin{center}
\begin{tabular}{|c|c|c|c|}
\hline
Parameter & SM & $m_{b'}$=800 GeV & $m_{b'}$=1200 GeV\\
\hline
$\theta_{12}$ &$0.2273 \pm 0.0007$ &$0.2271\pm 0.0008$& $0.2270\pm0.0008$\\
$\theta_{13}$ &$0.0035\pm 0.0001$ &$0.0038  \pm 0.0001$&$0.0038\pm0.0001$ \\
$\theta_{23}$ & $0.0397\pm 0.0007$& $0.0391\pm 0.0007$&$0.0391\pm 0.0007$  \\
$\delta_{13}$ & $1.10\pm 0.10$&$1.04\pm 0.08$&$1.04\pm 0.08$ \\
\hline
$\theta_{14}$ & -- & $0.0151 \pm 0.0154$& $ 0.0147 \pm 0.0149$\\
$\theta_{24}$ & -- &$0.0031 \pm 0.0039$& $0.0029\pm 0.0036$\\
$\theta_{34}$ &  -- & $0.0133 \pm 0.0130$&$0.0123\pm 0.0122$\\
$\delta_{14}$& -- & $0.11 \pm 0.22$&$0.11\pm 0.23$\\ 
$\delta_{24}$ & -- & $3.23 \pm 0.24$& $3.23 \pm 0.27$ \\
\hline
$\chi^2/d.o.f.$ & $82.42/60$ & $70.99/63$& $70.96/63$\\
\hline
\end{tabular}
\caption{The results of the fit to the parameters of CKM and ZFCNC.}
\label{table:parameters}
\end{center}
\end{table}

\begin{table}
\begin{center}
\begin{tabular}{|c|c|c|c|}
\hline
Qunatity & SM & $m_{b'}$= 800 GeV & $m_{b'}$= 1200 GeV \\
\hline
$|V_{ud}|$ &$0.9743 \pm 0.0002$& $0.9742\pm 0.0003$&$0.9742\pm 0.0003$ \\
$|V_{us}|$ &$0.225 \pm 0.001$ & $0.225 \pm 0.001$&$0.225 \pm 0.001$\\
$|V_{ub}|$ &  $(3.50 \pm 0.10) \times 10^{-3}$& $(3.80 \pm 0.10)\times 10^{-3} $ &$(3.80 \pm 0.10)\times 10^{-3} $\\
$|V_{ub'}|$ & -- & $ 0.0151\pm 0.0154$&$0.0147\pm 0.0149$\\
$|V_{cd}|$ & $0.225 \pm 0.001$&$0.225 \pm 0.001$&$0.2249 \pm 0.0008$ \\
$|V_{cs}|$ &$0.9735 \pm 0.0002$ & $0.9736 \pm 0.0002$&$0.9736 \pm 0.0002$\\
$|V_{cb}|$ &$0.040\pm 0.001$ &$0.0391 \pm 0.0007$&$0.0391 \pm 0.0007$ \\
$|V_{cb'}|$ & --  &  $0.0031\pm 0.0039 $ &$0.0029 \pm 0.0036$\\
$|V_{td}|$ &$0.0080\pm 0.0004$& $0.0074\pm 0.0004$&$0.0075\pm 0.0004$\\
$|V_{ts}|$ &$0.039\pm 0.001$& $0.0385 \pm 0.0007$& $0.0385 \pm 0.0007$\\
$|V_{tb}|$ &1 & $0.9991 \pm 0.0002$&$0.9991 \pm 0.0002$\\
$|V_{tb'}|$ & -- & $0.0133 \pm 0.0130$&$0.0123 \pm 0.0122$\\
\hline
\end{tabular}
\caption{Magnitudes of the $3\times 4$ CKM elements obtained from the fit. }
\label{table:ckm}
\end{center}
\end{table}

The magnitude of elements of the 3 $\times$ 4 quark mixing matrix, 
obtained by using the fit values presented in Table~\ref{table:parameters}, 
are given in Table~\ref{table:ckm}. Clearly all new elements of the 
quark mixing matrix are  consistent with zero. Furthermore, the 3$\sigma$ upper bound on the new CKM elements 
$V_{ub'}$,  $V_{cb'}$ and $V_{tb'}$ are $0.07$, $0.02$ and $0.06$, respectively indicating that the mixing of the $b'$
quark to the other three is very small. 

It is obvious from Table~\ref{table:ckm} that the values of CKM elements $V_{td}$ and $V_{ts}$ in ZFCNC model remains almost the same
as compared to their SM predictions. However, the allowed range of $V_{ub}$ gets slightly inflated. 
Because of this, the measured and predicted values of branching ratio of
$B\to \tau \,\bar{\nu}$  are in better agreement with each other in ZFCNC model in comparison to SM. This can be seen by comparing the 
$\chi^2_{B \to \tau \,\nu}$ contribution to the total $\chi^2_{\rm min}$ in ZFCNC model with that of SM. In SM, $\chi^2_{B \to \tau \,\nu}= 2.47$ which reduces to $0.91$ in the ZFCNC model indicating an improvement over the SM value.

 The $s\to d$, $b\to d$, and $b\to s$ transitions, which are the relevant ones for $K$ and $B$ decays, get contributions from terms involving the SM bilinears $\lambda^{i}_{jk} \equiv V_{ij}^{\ast} V_{ik}$ ($i \in \{u, c, t\}$ and $j,k \in \{d,s,b \}$)  and the new physics couplings $U_{jk}$ which are expressed in terms of $\lambda^{4}_{jk}$ ($U_{jk} = -V^*_{4j}V_{4k}=-\lambda^{4}_{jk}$). The values of the SM bilinears do not get much affected by the addition of the $b'$ quark. This is due to the fact that the SM CKM parameters  remains almost  unaffected. The allowed values of ZFCNC couplings $U_{sd}$, $U_{db}$ and $U_{sb}$ are 
given in Table~\ref{table:coupling}. It can be seen that there are large errors on them. For example, the new physics coupling relevant for rare $K$ decays, $U_{ds}$, is obtained to be $(0.27 \pm 5.89)\times 10^{-5}$. Although the best fit value is $2.7 \times 10^{-6}$ indicating tight constraint, due to large errors the 1$\sigma$ upper limit gets inflated upto $6.16\times 10^{-5}$. 
This is because these couplings are determined using the complicated functions of the nine CKM4 parameters with highly-correlated errors (by adding all errors in quadrature). 

The fit indicates that $|U_{sb}|<< |V^{*}_{ts}V_{tb}|$. Therefore 
new physics contribution in $b \to s$ sector is expected to be small in ZFCNC model.
 This can be seen, for example, from the study of observable 
$P'_5$ in bin $[4.3-8.68]$ GeV$^2$. The discrepancy between the experimental measurement and 
the SM prediction of $P'_5$ in this bin is around the 4$\sigma$ level. 
In the SM fit, $\chi^2_{P'_5}$ contribution 
to the total $\chi^2_{\rm min}$ is 16.94 indicating the disagreement between the
experimental measurement and SM prediction. In ZFCNC fit, we find $\chi^2_{P'_5}=17.00$, which is almost the same as in the SM. 

The like-sign dimuon charge asymmetry in the $B$ system, $A^b_{SL}$,  
receives contribution from both $b \to s$ and $b \to d$ sector.
The experimental measurement of $A^b_{SL}$  is 3$\sigma$  away from the SM prediction. 
In the SM fit, $\chi^2_{A^b_{SL}}$ contribution to the total $\chi^2_{\rm min}$ is  7.73
indicating this discrepancy. In ZFCNC fit, we find $\chi^2_{A^b_{SL}}=6.68$, indicating only a slight improvement over the SM value.

\begin{table}
\begin{center}
\begin{tabular}{|c|c|c|c|}
\hline
Quantity & $m_{b'}$= 800 GeV & $m_{b'}$= 1200 GeV\\
\hline
$|U_{ds}|$  & $(0.27 \pm 5.89)\times 10^{-5}$&$(0.15 \pm 1.91)\times 10^{-5}$ \\
$|U_{db}|$  & $(2.05 \pm 2.84)\times 10^{-4}$& $(1.84 \pm 2.56)\times 10^{-4}$\\
$|U_{sb}|$  &$(0.23\pm 5.17)\times 10^{-5}$& $(0.12\pm 1.51)\times 10^{-5}$\\
\hline
\end{tabular}
\caption{Magnitude  of ZFCNC couplings.}
\label{table:coupling}
\end{center}
\end{table}

\section{Predictions for other observables}
\label{pred}
We now turn on to predict some of the observables which are expected to deviate from their SM predictions due to addition of a $b'$ quark. In ZFCNC model, the flavor changing neutral current transitions  occur at tree level in the down sector whereas in the up sector, they occur at loop level. Hence the flavor signatures of ZFCNC model are expected to be coming from observables in the $K$ and $B$ sector. 
Given the tight constraints on new physics couplings obtained here, it will be interesting to see whether large deviations from SM is still allowed for some of the observables.

\subsection{\bf \boldmath Branching fraction of $K_L \to \pi^0 \nu \bar{\nu}$} 
The branching fraction of $K_L \to \pi^0 \nu \bar{\nu}$, which is governed by CP violation, in ZFCNC model is \cite{AguilarSaavedra:2002kr} 
\begin{eqnarray}
{\cal B}(K_L\to \pi^0\nu\bar{\nu}) &=& r_{K_L}\frac{\tau_{K_L}}{\tau_{K^+}}
\frac{\alpha^2  {\cal B}(K^+\to \pi^0 e^+ \bar{\nu}) }{2\pi^2 \sin^4\theta_W |V_{us}|^2} \times  \sum_{\rm l = e,\mu,\tau}\Bigl[X_{NL}^l\,Im(\lambda_{ds}^c) \nonumber\\
&&+~ \eta_t^{X}X_0(x_t)Im(\lambda_{ds}^{t})-\frac{\pi^2 Im(U_{ds})}{\sqrt{2}G_FM_W\,^2}\Bigr]^2\,,
\label{brkpi0}
\end{eqnarray}
where $r_{K_L}$ is the isospin breaking correction in relating $K_L\to \pi^0\nu\bar{\nu}$ to $K^+\to \pi^0 e^+ \bar{\nu}$.
$\eta_{X}$ is the NLO QCD correction, its value is estimated to be $0.994$\cite{Buras:1997fb}. The function $X_0(x_t)(x_t = m_t^2/M_W^2)$ is given by
$$X_0(x_t) = \frac{x_t}{8}\Bigl[-\frac{2+x_t}{1-x_t}+\frac{3x_t-6}{(1-x_t)^2}ln\,x_t\Bigr].$$
The SM prediction for the branching ratio of $K_L \to \pi^{0} \nu \bar{\nu}$ is given by {\cite{Buras:2006gb,Brod:2008ss}}
\begin{equation}
{\cal B}(K_L \to \pi^{0} \nu \bar{\nu}) = (2.27 \pm 0.28)\times 10^{-11}.
\end{equation}
The present experimental upper bound on its branching ratio is $2.6\times 10^{-8}$ at $90\%$ C.L. \cite{Ahn:2009gb}, which is about three orders of magnitude above its SM prediction.

 Using Table~\ref{table:parameters}, we get ${\rm Im}(U_{ds})=(1.83\pm 16.40)\times 10^{-6}$, for $m_{b'}$=800 GeV, which gives ${\cal B}(K_L\to \pi^0\nu\bar{\nu}) = (0.03 \pm 4.29)\times 10^{-11}$. At 2$\sigma$, ${\cal B}(K_L\to \pi^0\nu\bar{\nu})\leq 8.61 \times 10^{-11}$, indicating  that large enhancement in ${\cal B}(K_L\to \pi^0\nu\bar{\nu})$ above its SM value is not possible in the ZFCNC model.

\subsection{ \bf \boldmath Branching fraction of $B \to X_s \nu \bar{\nu}$}
In the SM, the decay $B\to X_s \nu \bar{\nu}$ is dominated by the
$Z^0$ penguin and box diagrams involving top-quark exchange, and is
theoretically clean.
The branching fraction for $B\to X_s \nu \bar{\nu}$ in ZFCNC model is given by 

\beq
{\cal B}(B\to X_s \nu \bar{\nu}) = \frac{\alpha^2}{2\pi^4\sin^4\theta_W}{\cal B}(B\to X_c e \bar{\nu})\frac{\bar{\eta}|V_{ts}^{*} V_{tb}X_0^{'}(x_t)|^2}{|V_{cb}|^2 f(\hat{m}_c)\kappa(\hat{m}_c)} 
\eeq
where $X_0^{'}(x_t)$ is the structure function in ZFCNC model given by \cite{Alok:2012xm}
$$X_0^{'}(x_t) = X_{0}(x_t) + \Bigl(\frac{\pi\sin^2\theta_W}{\alpha V_{ts}^{*}V_{tb}}U_{sb}\Bigr)$$.
The factor $ \bar \eta \approx 0.83$ represents the QCD correction to
the matrix element of the $b\to s\nu{\bar{\nu}}$ transition due to
virtual and bremsstrahlung contributions, $f(\hat{m}_c)$ is the
phase-space factor in ${\cal B}(B\to X_c e\bar{\nu})$, and
$\kappa(\hat{m}_c)$ is the 1-loop QCD correction factor.  The SM
prediction for ${\cal B}(B \to X_s \nu \bar{\nu})$ is $(2.28\pm 0.19)
\times 10^{-5}$, while in the ZFCNC model,
 this branching ratio is predicted to 
be  $(2.27 \pm 0.55)\times 10^{-5}$ for $m_{b'}$=800 GeV. Therefore a large
enhancement in the branching fraction of $B\to X_s \nu \bar{\nu}$ is
not allowed.

\subsection{ \bf \boldmath Direct $CP$ asymmetry in $B \to (K,\,K^*)\,\mu^+\,\mu^-$}
In the SM, the direct $CP$ asymmetry in the $b \to s\,\mu^+\,\mu^-$ modes is expected
to be very small. Indeed, in SM the Wilson coefficients $C_7$ and $C_{10}$ are real,
while the Wilson coefficient $C^{\rm eff}_9$ becomes only slightly
complex due to the on-shell parts of the
$u\bar{u}$ and $c\bar{c}$ loops, which are proportional to 
$V_{ub}^*V_{us}$ and $V_{cb}^*V_{cs}$, respectively. 
This complex nature of $C^{\rm eff}_9$ is the only source of
$CP$ asymmetry in the SM. 

Here we consider direct $CP$ asymmetry in the branching ratio of $B \to (K,\,K^*)\,\mu^+\,\mu^-$
which is defined as
\begin{equation}
A_{\rm CP}=\frac{B\left(\bar{B} \to (\bar{K},\,\bar{K^*})\,\mu^+\,\mu^-\right) - B\left(B \to (K,\,K^*)\,\mu^+\,\mu^-\right)} 
{B\left(\bar{B} \to (\bar{K},\,\bar{K^*})\,\mu^+\,\mu^-\right) + B\left(B \to (K,\,K^*)\,\mu^+\,\mu^-\right)}\;,
\end{equation}
where $B$  represent the branching ratios of the given mode. 
Within the SM $A_{\rm CP}\sim{\cal O}(10^{-3})$ \cite{Altmannshofer:2008dz}. 
The interference between the $C^{\rm eff}_9$ term and the new physics coupling 
terms can enhance $A_{\rm CP}$ up to $\pm 0.15$ \cite{Alok:2011gv}. Due to 
large errors, the present measurements for these modes are consistent with 
the SM prediction of small $CP$ asymmetry \cite{Aaij:2014bsa}.

Due to the extended quark mixing matrix, there are additional $CP$ violating phases in the ZFCNC model. Therefore one expects to have large enhancement in the $CP$ asymmetry. However due to tight constraints on the new physics couplings, the enhancement can only be up to 3-4 times that of the SM which could be too small to be observed at the LHC with current precision. 

\subsection{ \bf \boldmath Deviations in $Wtb$ coupling}
Due to the non unitarity of the quark mixing matrix, one can expect 
deviation of $|V_{tb}|$ from unity in this model. In the SM, $|V_{tb}|$ is 
determined using the unitarity condition. The direct determination of $|V_{tb}|$ 
without assuming unitarity is possible from the single top-quark-production cross section.
The CDF and D$0$ measuremnt gives $|V_{tb}|=1.03\pm0.06$ \cite{tevatron-vtb}
whereas the LHC measuremnts gives $|V_{tb}|=1.03\pm0.05$ \cite{lhc-vtb}. 
Although the present measurements have large errors, they do 
not rule out large deviations of $|V_{tb}|$ from unity.
We find  $|V_{tb}| = 0.9991 \pm 0.0002$. Thus, at $3\sigma$, we have $|V_{tb}| \ge 0.99$.  
Therefore this model cannot account for any large deviation of $|V_{tb}|$ from unity.
The possible deviation in the $Wtb$ coupling, i.e.,  $|V_{tb}|-1$ is $0.0009\pm 0.0002 $.
Thus at $3\sigma$, deviations in bottom coupling to $W$ can be only be up to $0.2 \%$ which is too small to 
be observed in the single top production at the LHC \cite{lhc-vtb}.

\subsection{\bf \boldmath Deviations of the bottom couplings to Higgs boson}
The Lagrangian of the SM bottom quark modified by the mixing with 
vector-singlet $b'$ quark is given by \cite{Aguilar-Saavedra:2013qpa}
\beq
{\mathcal{L}}_{H} = - \frac{g\,m_b}{2M_W} X_{bb}\,\bar{t}t\,H,
\eeq
where $X_{bb} = 1-|V_{4b}|^2$. Hence within the SM, $X_{bb} = 1$. 
Therefore, possible deviations of the 
bottom quark couplings to the Higgs boson is given by
\beq
\Delta X_{bb} = X_{bb} - (X_{bb})^{SM} = X_{bb} - 1 = -|V_{4b}|^2.
\eeq
Using our fit results, we get
\beq
\Delta X_{bb} =  -(0.17\pm 0.34)\times 10^{-3}.
\eeq
Thus at $3\sigma$, the possible deviation in the Higgs Yukawa 
coupling is  $<0.2$\% which is again too small to be observed at LHC with the current precision.

\section{Conclusions}
\label{concl}
In this paper we consider the minimal extension of SM by addition 
of an isosinglet, vector like
down-type quark $b'$. Using input from many flavor-physics processes, we
perform a $\chi^2$ fit to constrain the elements of the $3\times 4$
quark-mixing matrix and the ZFCNC couplings. The fit takes into account both
experimental errors and theoretical uncertainties. 

We conclude the following:
\begin{itemize}

\item The best-fit values of all three new real parameters of the CKM4
matrix are consistent with zero. 

\item The values of $V_{ts}$ and $V_{td}$ in this model are close to their SM predictions.

\item   The mixing of the $b'$ quark with the other three is constrained to be 
$|V_{ub'}|< 0.07$, $|V_{cb'}|< 0.02$, and $|V_{tb'}|< 0.06$ at $3\sigma$.

\item The tree level ZFCNC couplings are constrained to be small. At  $3\sigma$, 
$U_{ds} \leq 1.8 \times 10^{-4}$, $U_{db}\leq 1.1 \times 10^{-3}$ and $U_{sb}\leq 1.6 \times 10^{-4}$.

\item Large enhancement in the branching ratio of $K_L\to \pi^0\nu\bar{\nu}$ and $B \to X_s \nu \bar{\nu}$ is not allowed.

\item Large enhancement in direct $CP$ asymmetry in $B \to (K,\,K^*)\,\mu^+\,\mu^-$ is not allowed.

\item The deviations in $Wtb$ coupling as well as SM bottom quark coupling to Higgs is too small to be measured at the LHC with current precision.

\end{itemize}

Therefore we observe that the current flavor physics data puts tight constraints on ZFCNC model. The possibility of detectable new physics signals in most of the flavor physics observables is ruled out for this model. 

{\em Acknowledgments.\textemdash}
 The work of AKA and SB is supported by CSIR, Government of India, grant no: 03(1255)/12/EMR-II.

\end{document}